\documentclass[12pt]{iopart}          
\usepackage{url}

\RequirePackage[T1]{fontenc}

\RequirePackage{multirow}
\RequirePackage{graphicx}
\RequirePackage{mathptmx}      
\RequirePackage{flushend}
\RequirePackage[numbers,sort&compress]{natbib}
\RequirePackage[colorlinks,citecolor=blue,urlcolor=blue,linkcolor=blue]{hyperref}

\usepackage{graphicx}  
\usepackage{dcolumn}   
\usepackage{bm}        
\usepackage{amssymb}   
\usepackage{feynmf}    
\usepackage{hyperref}  %
\usepackage{slashed}
\usepackage{color}
\usepackage{array}
\usepackage{caption}
\usepackage{lineno}
\usepackage{subcaption}
\expandafter\let\csname equation*\endcsname\relax
\expandafter\let\csname endequation*\endcsname\relax
\usepackage{amsmath}
\usepackage{booktabs}
\usepackage{multirow}
\usepackage{subcaption}
\usepackage{wrapfig}
\usepackage{dblfloatfix}

\newsavebox\mybox

\begin{document}

\title{Set-Conditional Set Generation for Particle Physics}

\author{Nathalie Soybelman$^1$, Nilotpal Kakati$^1$, Lukas Heinrich$^2$, Francesco Armando Di Bello$^3$, Etienne Dreyer$^1$, Sanmay Ganguly$^4$, Eilam Gross$^1$, Marumi Kado$^{5,6}$, and Jonathan Shlomi$^1$}
\address{$^1$ Weizmann Institute of Science}
\address{$^2$ Technical University of Munich}
\address{$^3$ INFN and University of Genova}
\address{$^4$ ICEPP, University of Tokyo}
\address{$^5$ INFN and Sapienza University of Rome}
\address{$^6$ Max Planck Institute for Physics}

\ead{nathalie.soybelman@weizmann.ac.il, nilotpal.kakati@weizmann.ac.il, and l.heinrich@tum.de}
\vspace{10pt}
\begin{indented}
\item[]November 2022
\end{indented}

\begin{abstract}
The simulation of particle physics data is a fundamental but computationally intensive ingredient for physics analysis at the Large Hadron Collider, where observational set-valued data is generated conditional on a set of incoming particles. To accelerate this task, we present a novel generative model based on a graph neural network and slot-attention components, which exceeds the performance of pre-existing baselines.
\end{abstract}
\vspace{1pc}
\vspace{1pc}

\section{Introduction}
\begin{figure}
    \centering
    \includegraphics[width=\textwidth]{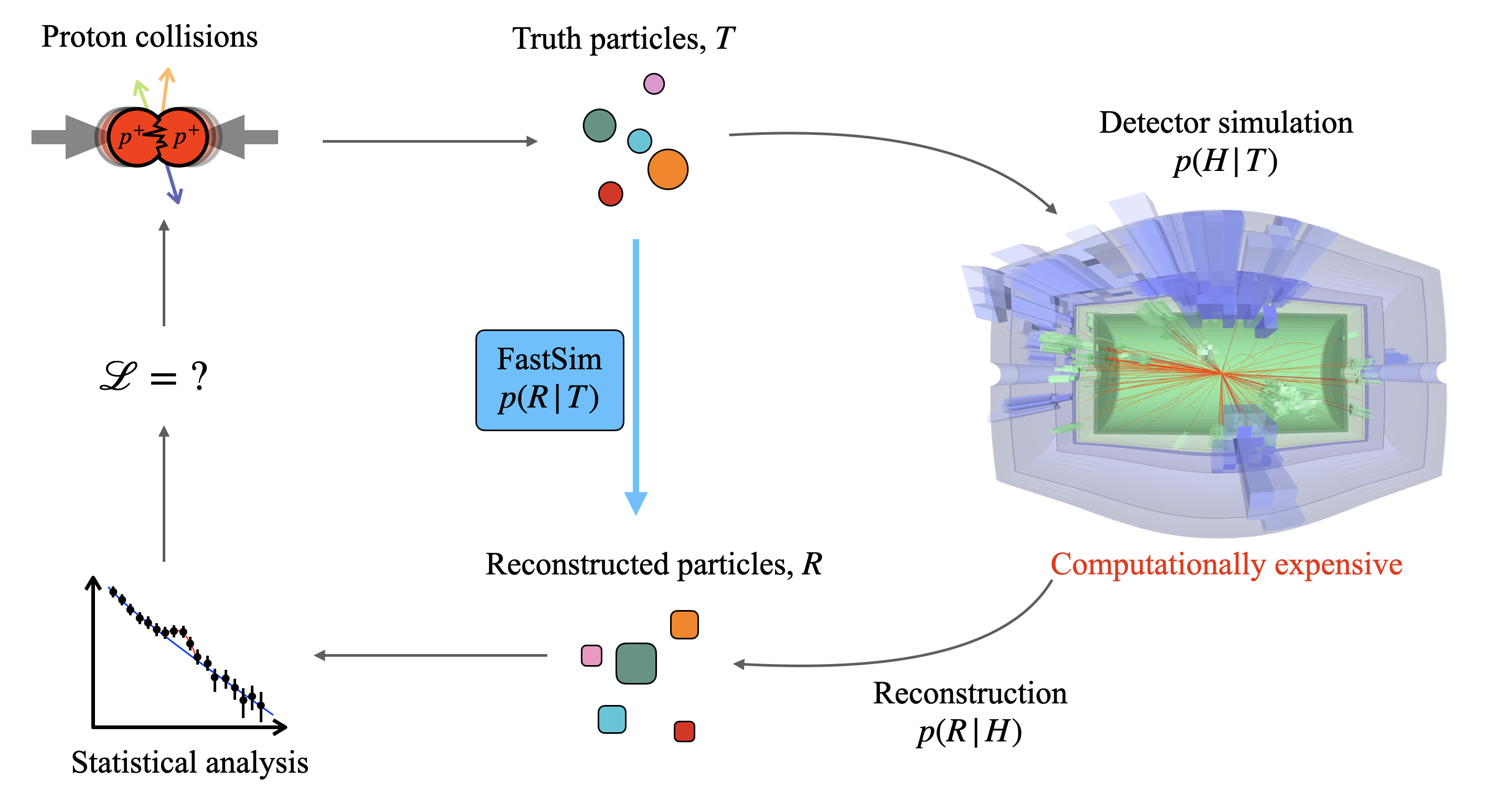}
    \caption{Classical simulation pipeline in particle physics: collision simulation generates truth particles $T$, detector simulation produces high-dimensional signals $H$, and reconstruction algorithm recovers truth input from detector readout as reconstructed particles $R$. Proposed fast simulation skips detector simulation and reconstruction, directly obtaining reconstructed particles from the truth.}
    \label{fig:intro}
\end{figure}
The most computationally expensive and challenging tasks in high-energy physics at collider experiments are the simulation and reconstruction of collision events. We display the key steps of the simulation pipeline in fig.~\ref{fig:intro}. The collision simulation includes modeling of the underlying physics process in the proton-proton collision (e.g., $t\bar{t}$ event) based on its matrix element, showering, and hadronization~\cite{rev_eventgen}. This gives us a variable-sized \emph{set} of ``truth'' particles $T = \{t_i| i = 1\dots N_T\}$. Subsequently, the interaction of the particles with the detector needs to be modeled. Simulation tools such as \textsc{Geant4}~\cite{geant} use microphysical models to simulate the detailed stochastic interactions with the detector material producing a set of signals (``hits'') $H$ in the read-out sensors of the detector. This simulation implicitly corresponds to an underlying distribution $p_\mathrm{sim}(H|T)$.
Modern detectors have up to a hundred million such sensors, so the high-dimensional space of hits is inconvenient for physics analysis. ``Reconstruction'' is a deterministic inference algorithm $R(H)$ that attempts to recover approximately the set-valued latent input $T$ to present physicists with an interpretable and low-dimensional summary of the hit data as a set of \emph{reconstructed} particles $R = \{r_i|\;i=1\dots N_R\}$ that aims to approximate $T$. Typically, physicists do not directly interact with the hit-level data but only with the effective set-valued model 
\begin{equation}
    R \sim p(R|T) = \int \mathrm{d}H\, \delta(R(H)-R)p_\mathrm{sim}(H|T).
\end{equation} Due to the high computational cost of both simulation and reconstruction, there is considerable interest in exploiting generative machine learning to develop fast surrogates for this effective model. This work explores the possibility of training an end-to-end surrogate $R\sim q_\theta(R|T)$ with learnable parameters $\theta$. We split the generative process into a cardinality prediction task $q_{\theta_1}(N_R|T)$ and a doubly conditional set generation task $q_{\theta_2}(R|N_R,T)$. The necessary permutation invariances implied by the set nature of $T$ and $R$ are enforced through inductive biases in the architecture. Our results exceed the performance of baseline models.
\subsection{Related Work}
\label{ssec:related}
There are two main approaches to approximate $p(R|T)$. In one approach, only $p_\mathrm{sim}(H|T)$ is replaced by a fast surrogate such as~\cite{Paganini:2017dwg,caloflow,graphcalo,DLCalo,EMshowerGAN}. Based on its output, a reconstruction algorithm may be used to produce reconstructed particle sets $R(H)$. New ML-based reconstruction algorithms have been developed in recent years, such as~\cite{CMSpflow,HGpflow,MLPF}, and can be applied for this task. This approach has two disadvantages: Firstly, the surrogate must correctly learn a high-dimensional generative model $H\sim p(H|T)$ only for $H$ to be further processed and reduced in dimension to form reconstructed events. Secondly, this approach incurs the full cost of the standard reconstruction, which is still significant. The second approach aims for a direct approximation of the lower-dimensional $p(R|T)$. Prior work simplifies the problem by first projecting the sets to fixed-sized feature vectors of the truth and reconstructed events $\mathbf{t} = f_t(T)$, $\mathbf{r} = f_r(R)$, and aims to learn a fast generative model $p(\mathbf{r}|\mathbf{t})$~\cite{ATLAS:2021pzo,Butter:2019cae,graphgan,ParticleGAN,MPGAN}. More literature can be found in~\cite{livingreview}. This approach is limited by the \emph{fixed} choice of $f_r,f_t$ and thus does not enable one to generate features outside of $f_r$. A fully general approach modeling $p(R|T)$ is only possible through a \emph{set-to-set} approach. Prior work has aimed at learning a fast surrogate using a set-valued variational autoencoder (VAE)~\cite{vae_fastsim_jets}, similar to the baseline we present in this work. The encoder yields a latent distribution $p(z|T)$, and a decoder implements a model of reconstructed events $p(R|z)$. 
While this model successfully reproduces the \emph{marginal} distributions (i.e. projections of $p(R) = \int\mathrm{d}T\;p(R|T)p(T)$), the authors note that ``[the algorithm] fails in faithfully describing the jet dynamics at constituents level'' and do not present non-marginal results. To the best of our knowledge, this is the first work presenting an extensive analysis of the conditional distribution of such a set-to-set model in a particle physics application.
\section{Dataset}
We demonstrate the set-to-set approach using a simplified ground truth model of $p(R|T)$ for charged elementary particles represented by their direction and momentum features ($p_\mathrm{trans.}, \eta, \phi$). 
In order to learn the cardinality and conditional set generation models $q(N|T)$ and $q(R|T,N)$, we generate a set of representative truth events. Then, multiple samples of the target distribution $\mathcal{D} = \{(R_{ji},T_j) \sim p(R_i|T_j)p(T_j)\}$ are generated for each truth event by simulating the detector response and subsequent reconstruction. Specifically, the detector response is simulated multiple times for a single truth event. The reconstruction algorithm is applied each time, resulting in multiple reconstructions $R_i$, referred to as \emph{replicas}, for the same truth event $T$.
The replica cardinalities $N$ then serve as labels for the supervised training of the cardinality prediction. At the same time, the conditional empirical distribution of replicas for a given truth event serves as training samples for the set generation task. 
The training, validation, and test data set consist of 2915, 500, and 3990 truth events, respectively. For training (evaluation), 25 (100)  independent reconstructions are generated per truth event. The data set is provided in~\cite{note:zenodo}. 
\subsection{Truth Event Generation}
In the present work, we focus on a localized reconstruction of particles within a single jet. Events are produced using \textsc{Pythia8}~\cite{pythia} to generate a single quark with momentum between 10 GeV and 200 GeV with initial direction randomly chosen in the ranges $|\eta| < 2.5$ and $|\phi| < \pi$. Following parton shower and hadronization, only stable charged particles with momentum above 1 GeV and $|\eta| < 3.0$ are selected. The set of charged particles inside the jets has an average cardinality of $N_\mathrm{ch} = 3.72$ with a maximum of 12 particles and a minimum of 1 particle per truth event. This study focuses on the feasibility of correctly modeling the detector resolution. As a toy model, we currently use smeared tracks as targets to reconstruct the known track smearing instead of a full particle flow reconstruction with an unknown smearing model. As such, we are forced to limit ourselves to charged particles.

\subsection{Tracking emulation and smearing}
Each charged particle in the generated truth event is assumed to be generated at the origin. The trajectory of the charged particle through the inner detector, i.e., track, is parameterized by five perigee parameters called $d_0$, $z_0$, $q/p$, $\theta$, $\phi$. Here $q$ and $p$ are the charge and magnitude of the three-momentum of the associated particle. $\theta$ and $\phi$ parameterize the unit vector along the momentum direction. $d_0$ and $z_0$ are associated with the track curvature.
The effect of reconstruction of charged particles is emulated by smearing the $q/p$, $\theta$, and $\phi$ values of their associated track. Each is independently varied by a Gaussian resolution model with width dependent on the transverse momentum of the charged particle. No particle-particle correlations and no correlations of track parameter uncertainties are considered.
Additionally, a deterministic model is used to introduce tracking inefficiency. Charged particles are dropped if they were produced far from the beamline in terms of transverse radius R (R > 75 mm for $|\eta|<1.5$ and R > 250 mm otherwise). This results in an efficiency of $95.3\pm0.8\,\%$ over the combined training, validation, and test dataset. This accounts for the difference between the cardinalities of the truth and target sets.
%
%
\section{Models and Training}
In this work, we compare a novel neural network architecture based on a graph neural network~\cite{graphs} and slot-attention~\cite{slotattention} to a baseline model in the form of a conditional variational auto-encoder. In the following, we briefly describe the architectures and training process for both models.
\subsection{Conditional VAE}
\savebox{\mybox}{\includegraphics[width=10cm]{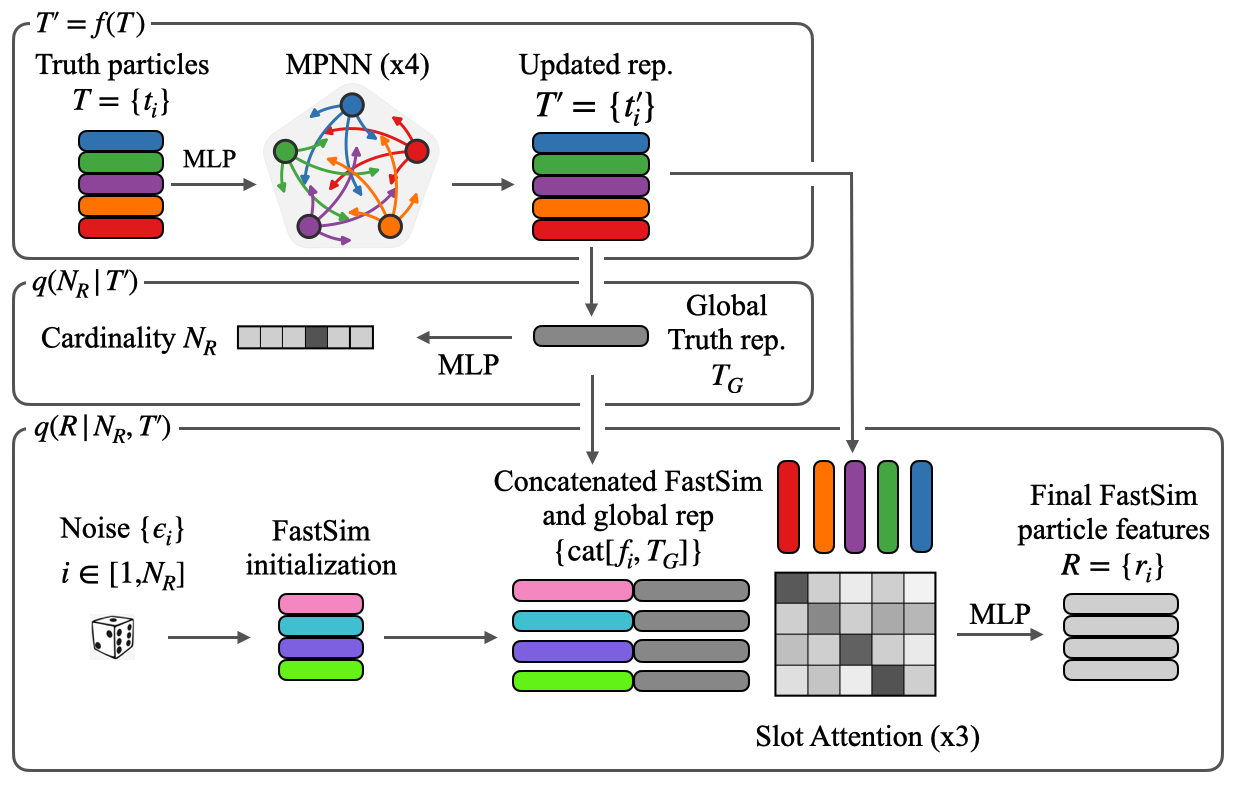}}
\begin{figure}[t]
    \centering
    \begin{subfigure}{0.31\textwidth}
        \centering
        \vbox to \ht\mybox{%
            \vfill
            \includegraphics[width=5cm]{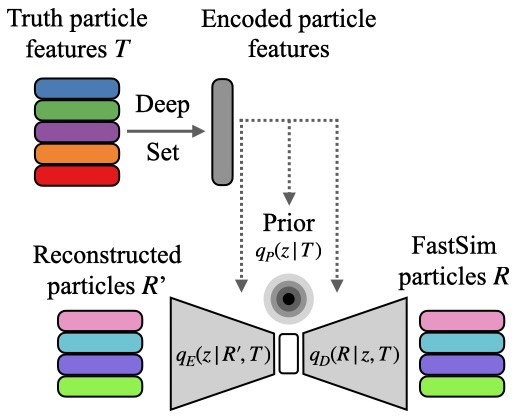}
            \vfill
        }
        \caption{cVAE}
        \label{fig:cvae_architecture}
    \end{subfigure}
    \begin{subfigure}{0.68\textwidth}
        \centering
        \usebox{\mybox}
        \caption{GNN+SA}
        \label{fig:gnn_architecture}
    \end{subfigure}
    \caption{Architectures of the two models}
    \vspace{-10pt}
\end{figure}
For the baseline results, we introduce a new benchmark model based on a conditional variational auto-encoder~\cite{cvae} to approximate the distribution $p(R|T)$ by a variational model $q_\phi$. The generated set will be referred to as $R$, and the target set to be encoded with $R'$. We extend the typical construction of variational auto-encoders~\cite{VAE} by conditioning the prior $q_{P}(z|T)$, an encoder $q_{E}(z|R',T)$ and a decoder $q_{D}(R|z,T)$ on the truth event $T$. As both $R$ and $T$ are sets, we encode them as Deep Sets~\cite{deepset} before passing them into the encoder and prior to ensure permutation invariance. 
The cVAE input is zero-padded to a maximum cardinality to handle the variable number of input particles. The output of the decoder network includes an additional presence variable to indicate whether the corresponding vector is to be considered a member of the output set. The threshold value for the presence variable was optimized through a grid search to 0.6. A sketch of the cVAE architecture is shown in Figure~\ref{fig:cvae_architecture}, and the network parameters are given in Table~\ref{tab:param} (left).
To generate reconstructed events $R$, a latent code is sampled from the conditional prior $z\sim q_P(z|T)$, which is then passed through the decoder to produce candidate output vectors. The presence of each particle is then sampled according to its indicator variable. 
\subsection{Graph Neural Network and Slot-Attention Model}
In addition to the cVAE, we present results on a novel model that uses a combination of graph neural networks and slot-attention layers. The general strategy is to, first, use a graph neural network with message passing $f_{\theta_0}$ to encode the truth particles $T=\{t_i\}$ into a high-dimensional vector representation $T'=\{t_i'\}$, then, predict the corresponding cardinality $N_R$ of the reconstructed event, and finally generate the reconstructed event $R$ at the predicted cardinality by transforming noise vectors into hidden representations of the reconstructed particles from which per-particle attributes such as $p_{trans.}$ can be projected out using simple multilayer-perceptron (MLP) networks as shown in Figure~\ref{fig:gnn_architecture}. The cardinality prediction corresponds to a model $q_{\theta_1}(N_R|T')$, whereas the feature prediction implements the model $q_{\theta_2}(R|N_R,T')$.
\subsubsection{Input Set Encoding and Cardinality Prediction}
\begin{table*}[t]
\centering
\begin{subtable}{0.43\textwidth}
\centering
    \begin{tabular}{l|r}
    \toprule
                \textbf{Network} &  \textbf{Parameters}\\
                \midrule
          DeepSet for $T$ & 300 000\\
          DeepSet for $R$ & 300 000\\
          Encoder $q_E(z|R',T)$ & 400 000\\
          Prior $q_P(z|T)$ & 360 000\\
          Decoder $q_D(R|z,T)$ & 380 000\\
         \midrule
         Total & 1 740 000 \\
         \bottomrule
    \end{tabular}
    \end{subtable}
\begin{subtable}{0.55\textwidth}
\centering
    \begin{tabular}{l|l|r}
    \toprule
                \textbf{Network} & \textbf{Subpart} & \textbf{Parameters}\\
                \midrule
           \multirow{2}{2cm}{$f_{\theta_0}$} &  MLP & 47 000\\
          & MPNN & 241 000 \\
          \midrule
          $q(N_R|T')$ & MLP & 12 000 \\
          \midrule
          \multirow{3}{2cm}{$q(R|N_R,T')$} & Embedding & 500 \\
          & Slot-Attention & 1 367 000 \\
          & MLP & 14 000\\
         \midrule
         Total & & 1 681 500 \\
         \bottomrule
    \end{tabular}
\end{subtable}
\caption{Network parameters for the cVAE (left) and SA+GNN (right) architectures.}
\label{tab:param}
\end{table*}
This embedding is constructed through a fully connected Graph Neural Network, with one node per truth particle. The truth particle attributes $\{p_{trans.},\, \eta,\, \phi\}$ are embedded using an MLP and attached as node features. Once initialized, the graph undergoes multiple rounds of message passing to produce a final set of input feature vectors $\{t_i'\}$. A graph-level permutation-invariant pooling operation provides an overall embedding $T_G$ of the input truth event $T$. Given the graph-level embedding of the truth event $T$, the categorical distribution of the output set cardinality is predicted using a simple MLP head. The embedding and cardinality prediction architecture is shown in the upper part of Figure~\ref{fig:gnn_architecture}. 
\subsubsection{Conditional Set Generation}
The model for the output set generation for a fixed cardinality $q_{\theta_2}(R|N_R,T')$ is designed as a generative model transforming a set of noise vectors $\{\epsilon_i\},\; i = 1...N_R$ into a set of reconstructed particle feature vectors $R = \{r_i\} = f_{\theta_2}(\{\epsilon_i\},\{t'_i\})$ conditioned on the truth event. The architecture thus encodes an \emph{implicit} model $q(R|N_R,T')$, from which reconstructed events can be sampled even if an explicit evaluation of the likelihood is not possible.
For the initialization, particle indices $\{i\}; i = 1...N_R$ are embedded, and random noise is added. This gives the set of initialized reconstructed particles $\{f_i\}$. Afterwards, the conditional model is provided with the global graph-level encoding of the event $T_G$ through a set of concatenated input vectors $\{\mathrm{cat}[f_i,T_G]\}$. The output generation proceeds through a Slot Attention layer, in which the vectors corresponding to the output particles are represented as randomly initialized slots that attend over the provided truth particles through multiple rounds of iterative refinement. As an attention mechanism, standard query $q$, key $k$, and value $v$ embedding are used to update the reconstructed particle representation. Here, key and value are MLPs that take the truth particle representation as input, while query takes the noise with the global representation: 
\begin{equation}
    k=F_k(t'_i,t_i),\quad v=F_v(t'_i,t_i),\quad\text{and}\quad q=F_q(f_i,T_G).
\end{equation}
We add a skip connection for the key and value and use the initial truth particle features with the updated ones. 
The attention matrix $A$ is calculated by taking the softmax of the dot-product of the key and query with a normalization
\begin{equation}
    A= \mathtt{Softmax}\left(\frac{1}{\sqrt{D}}k\cdot q^T\right)
\end{equation}
where $D$ is the output dimension of $F_k$ and $F_q$.
To update the reconstructed set representation, the value $v$ is multiplied with the attention matrix $A$, sent through a GRU cell, a layer normalization, an additional MLP, and added to the initial feature vector as described in~\cite{slotattention}. 
\begin{equation}
    f_i \longrightarrow f_i\;\; +\;\;\mathtt{MLP}\left( \mathtt{LayerNorm}\left(\mathtt{GRU}  \left(v*A,f_i\right)\right)\right) 
\end{equation}
In total, three rounds of slot attention updates are applied, each with independent $k$, $q$, and $v$. The output finally provides high-dimensional embeddings of the reconstructed particles from which attributes are projected using an element-wise MLP. The set generation architecture is sketched in the lower half of Figure~\ref{fig:gnn_architecture}, and the network parameters are given in Table~\ref{tab:param} (right). The slot-attention mechanism is permutation-equivariant; thus, if the initial noise model is permutation invariant, the implicit model learned during training will also be.
\subsection{Training}
Generalizing from the non-conditional VAE case, we train the cVAE on the negative evidence lower bound loss ($ELBO$) as averaged over both observed reconstructions $R$ and conditioning values $T$:
\begin{align}
    L &= -\mathbb{E}_{T,R} \mathbb{E}_{z\sim q_E(z|R,T)}\; \log \frac{q_D(R|z,T)q_P(z|T)}{q_E(z|R,T)}\\
      &= -\mathbb{E}_{T,R} \mathbb{E}_{z}\; \log q_D(R|z,T) + D_\mathrm{KL}(q_E(z|R,T)||q_P(z|T))\nonumber
\end{align}
%
\noindent As shown above, the ELBO loss for a given $(R,T)$ pair can be decomposed into two components: the reconstruction loss $\mathbb{E}_{z\sim q_E(z|R,T)} \log q_D(R|z,T)$ and a regularizing term comparing the (now truth-dependent) prior $q_P(z|T)$ to the per-instance posterior $q_E(z|R,T)$ distribution through the Kullback-Leibler divergence. The reconstruction loss is taken to be the sum of distances in feature space $(p_{trans.}, \eta, \phi)$ after a particle-by-particle assignment through the Hungarian Algorithm~\cite{HUNGARIAN}. The KL divergence can be computed in closed form as both the posterior  $q_E(z|R,T)$ and the prior $q_P(z|T)$ are taken to be multivariate normal distributions with a diagonal covariance matrix $\Sigma$. Their respective distribution parameters $\mu$ and $\log \Sigma$ are computed through MLPs as a function of their respective conditioning values. We train for 500 epochs (7 hours) using the ADAM optimizer~\cite{kingma2017adam} with a learning rate of $5\cdot10^{-4}$ on a 24564MiB GPU
(NVIDIA RTX A5000).
\newline
\begin{wrapfigure}[10]{r}{0.36\textwidth}
    \centering
    \vspace{-30pt}
    \includegraphics[width=0.34\textwidth]{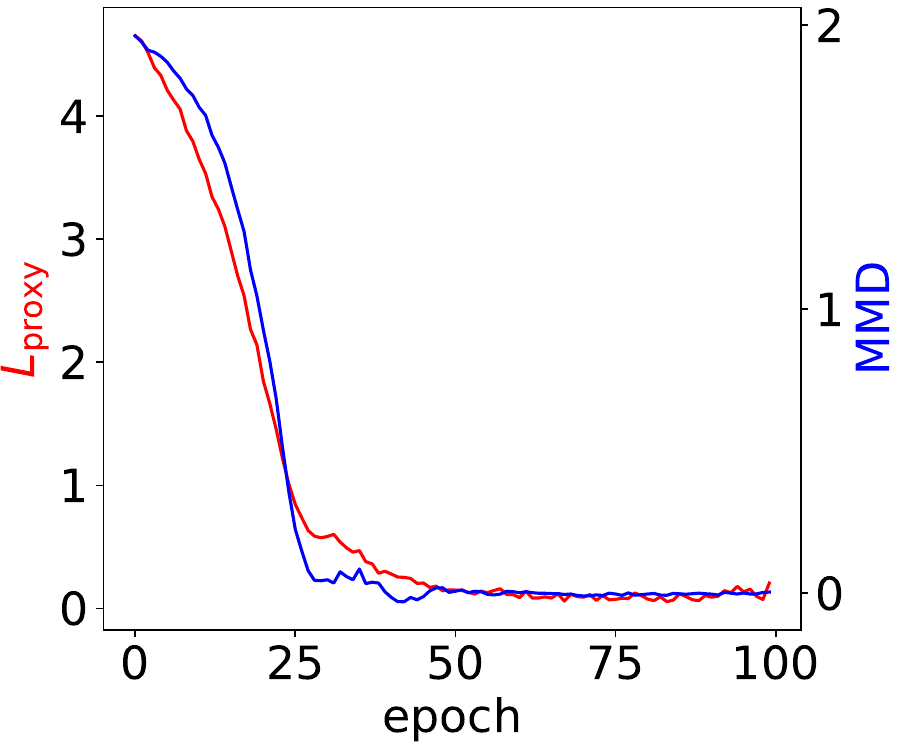}
    \caption{Evolution of MMD and $L_\mathrm{proxy}$ during training}
    \label{fig:losses}
\end{wrapfigure}
The GNN+SA model is trained on a combination of two tasks: cardinality prediction and set generation. The cardinality prediction is trained on a standard categorical cross-entropy loss $L_\mathrm{card.}$ in expectation over all truth events $T$. As the model does not provide a tractable likelihood $q_\phi(R|T,N)$, we formulate a sample-based similarity measure between the two distributions $q_\phi(R|T,N)$ and $p(R|T,N)$. A suitable metric is the maximum mean discrepancy ($\mathrm{MMD}^2$)~\cite{MMD} for which we use the Hungarian Cost kernel $k(x,x')$, which acts as a similarity measure between instances. We use the Hungarian Cost as the similarity measure.
\begin{equation}
    \mathrm{MMD}^2 = \mathbb{E}_{(x\sim p,x'\sim p)}[k(x,x')] + \mathbb{E}_{(x\sim q,x'\sim q)}[k(x,x')] - 2\mathbb{E}_{(x\sim q,x'\sim p)}[k(x,x')]
\end{equation}
While the MMD metric enjoys strong theoretical guarantees, such as vanishing when $p=q$, we observed empirically that training directly on it as a loss converges poorly. 
We thus use a heuristic proxy loss that facilitates training and empirically correlates well with the MMD, which we track during training as a metric. In Fig.~\ref{fig:losses}, we show that minimizing the proxy loss also minimizes the tracked MMD. Due to the cost of the MMD computation, the figure is shown for only one event.
In this proxy, we use the minimum kernel entry $L_\mathrm{proxy} = \mathrm{min}_{x_i,x'_j}k(x_i,x'_j)$, where $x_i$ is a member of the reference set and $x'_j$ is from the generative model sampling. It isThus, weer bound on the final $-2 \mathbb{E}_{p,q}[k(x,x')]$ term in the $\mathrm{MMD}$ definition. Effectively we perform a second Hungarian Algorithm matching the reconstructed replicas with the target once, in addition to the particle-by-particle assignment.
\noindent The total loss $L = L_\mathrm{card.}+L_\mathrm{proxy}$ is averaged over all truth events $T$.
Due to its expensive computation, we only train for 200 epochs (6 days) using the same optimizer, learning rate, and GPU as for the cVAE. The final models were selected based on their performance on the validation set. The code used in these studies is provided in~\cite{note:github}.
\section{Results}
\begin{figure*}[b!]
    \centering
    \begin{subfigure}[b]{0.325\textwidth}
        \centering
        \includegraphics[page=1,width=\textwidth]{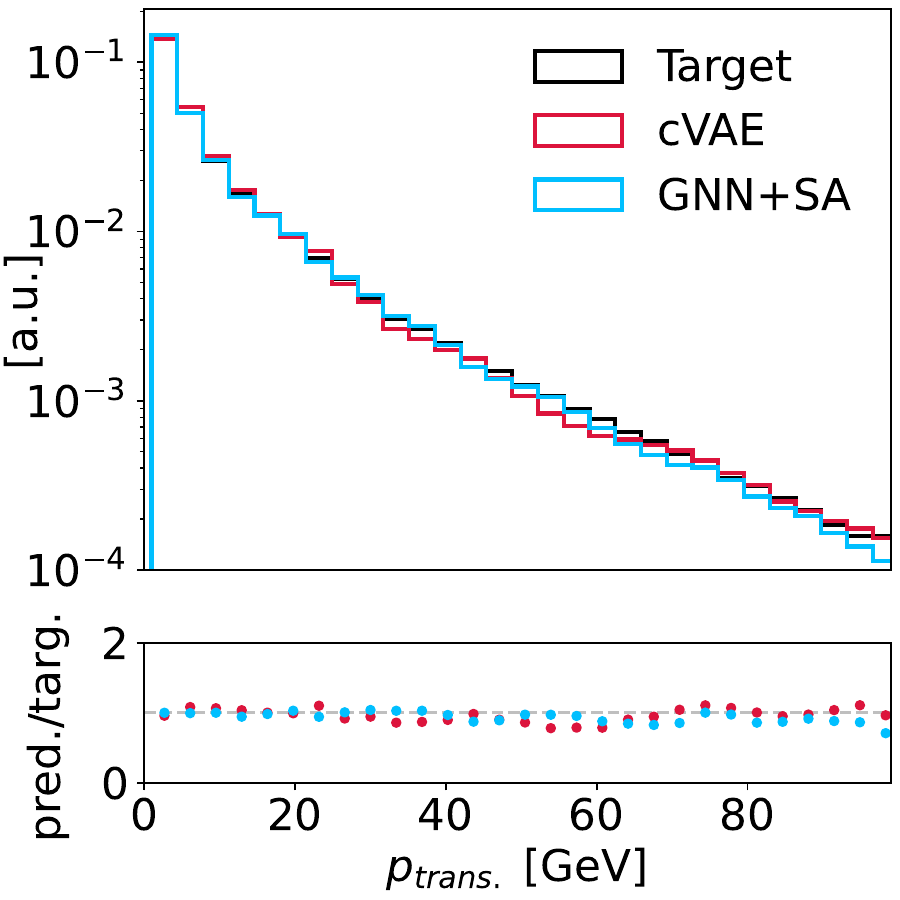}
    \end{subfigure}
    \begin{subfigure}[b]{0.325\textwidth}
        \centering
        \includegraphics[page=2,width=\textwidth]{fig/1d_histos_marginal.pdf}
    \end{subfigure}
    \begin{subfigure}[b]{0.325\textwidth}
        \centering
        \includegraphics[page=3,width=\textwidth]{fig/1d_histos_marginal.pdf}
    \end{subfigure}
    
       \caption{Marginal distribution of per particle observables $p_{trans.}$, $\eta$ and $\phi$}
    \label{fig:1d_pt}
      \vspace{-10pt}
\end{figure*}
\begin{figure*}[t!]
    \centering
    \begin{subfigure}[b]{0.48\textwidth}
    \includegraphics[width=\textwidth]{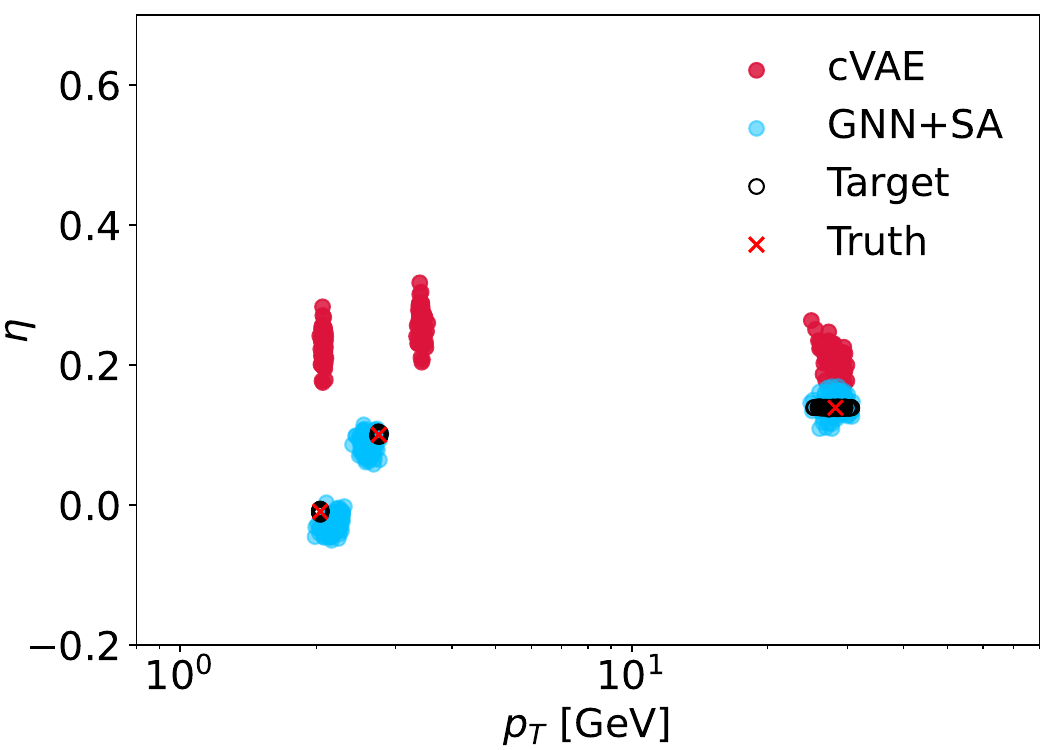}
    \caption{Event display}
    \label{fig:event_display}
    \end{subfigure}
    \hspace{12pt}
    \begin{subfigure}[b]{0.48\textwidth}
    \centering
    \includegraphics[width=\textwidth]{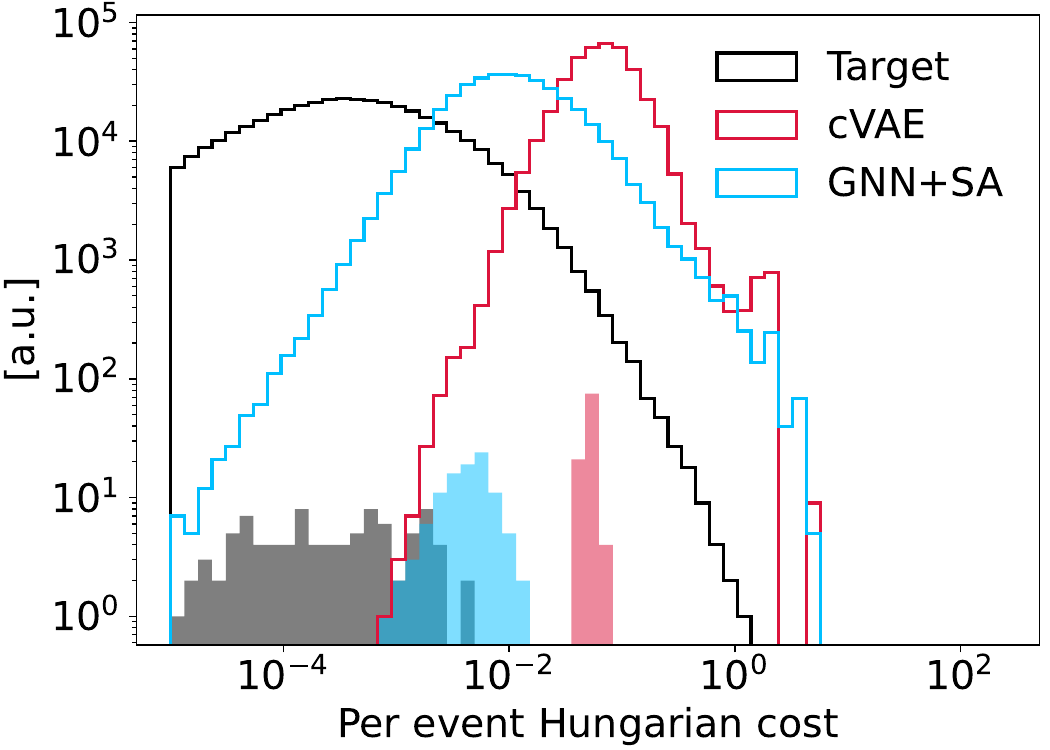}
    \caption{Hungarian cost distribution}
    \label{fig:hungarian_cost_distr}
    \end{subfigure}
    \caption{(a) The event display depicts the model predictions of the two networks alongside the ground truth and target. (b) The per-event set-to-set Hungarian cost, with the truth set as the reference, is shown for the entire test dataset. The filled histograms correspond to the loss distributions of the event in (a), which was chosen as a representative due to its cost distributions being near the peaks of the overall distributions.}
    \vspace{-10pt}
\end{figure*}

We present results for the conditional generation of reconstructed events of charged particles in terms of per-particle features and collective per-event set-level features. As the ground truth model has a deterministic relationship between the truth event $T$ and output cardinality $N_R$, we can assess the correctness of the model $p(N|T)$ by comparing the accuracy of the cardinality prediction with the ground-truth cardinality. Table~\ref{tab:metrics} shows that both models perform similarly after tuning cVAE hyper-parameters via grid-search. We also observe that marginal distributions are comparably reproduced by both models, confirming prior work in this area.
Figure~\ref{fig:1d_pt} shows the per-particle momentum as an example. Differences in the two models emerge when studying projections of the \emph{conditional} distributions $p(R|T)$. We present projected distributions $p(\mathbf{r}|\mathbf{t}) = p(f_r(R) = \mathbf{r}|f_t(T) = \mathbf{t})$, where $f_t(\cdot),f_r(\cdot)$ extract feature vectors on $T$ and $R$ respectively.
\begin{figure*}[t!]
    \centering
    \begin{subfigure}[b]{0.325\textwidth}
        \centering
        \includegraphics[width=\textwidth]{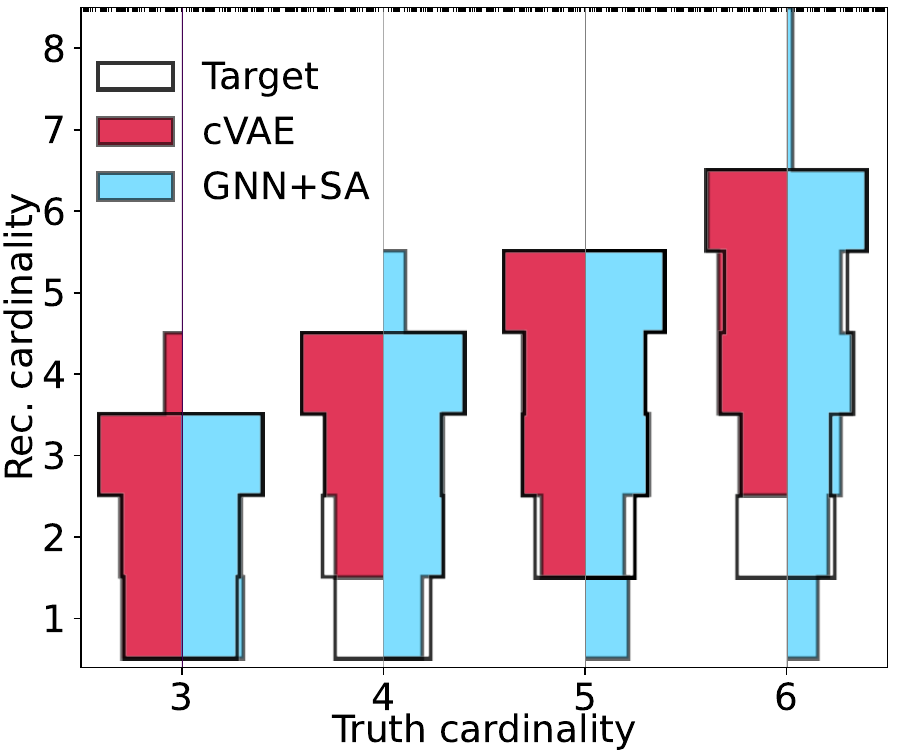}
        \caption{$p\big(N_R\big|N_T\big)$}
        \label{fig:conditional_cardinality}
    \end{subfigure}
    \begin{subfigure}[b]{0.325\textwidth}
        \centering
        \includegraphics[width=\textwidth]{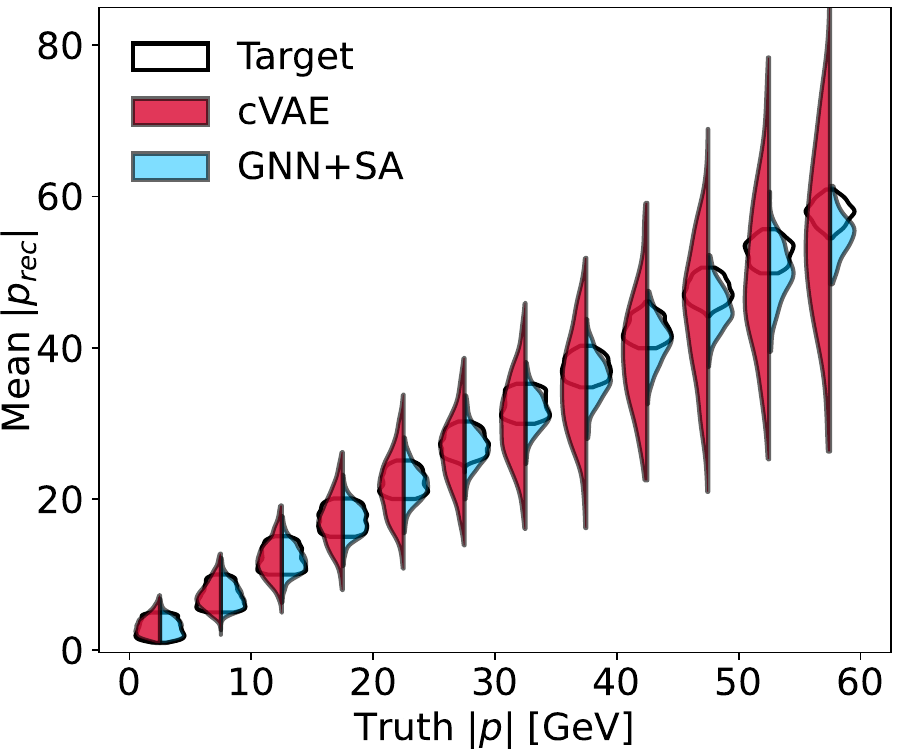}
        \caption{$p\big(\overline{(|p|_R)}\big||p|_T\big)$}
        \label{fig:conditional_mean}
    \end{subfigure}
    \begin{subfigure}[b]{0.325\textwidth}
        \centering
        \includegraphics[width=\textwidth]{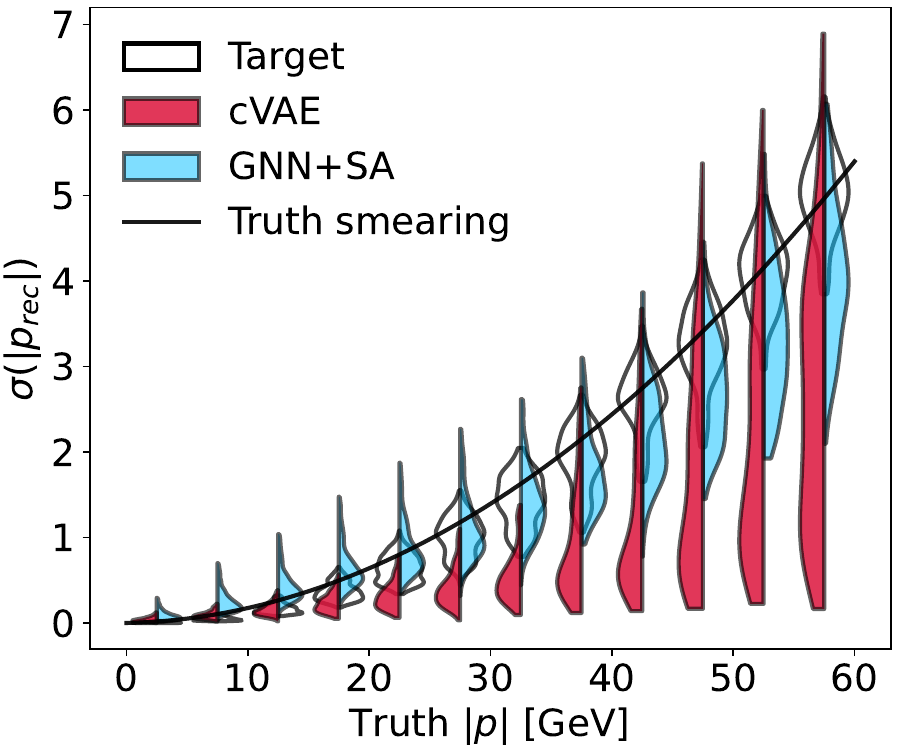}
        \caption{$p\big(\sigma(|p|_R)\big||p|_T\big)$}
        \label{fig:conditional_var}
    \end{subfigure}
    
       \caption{(a) Conditional Cardinality Distribution and (b and c) Distribution of reconstructed particle momentum means and their variances }
       \label{fig:conditional_distributions}
\end{figure*}
\begin{table*}[b]
\centering
    \begin{tabular}{c|c|c|c}
    \toprule
                &  Accuracy $q(N_R|T)$ [\%] & $\mathrm{MMD}^2\; q(R|N,T)$    & Hungarian Cost $C(R,T)$ $|\overline{C}_q - \overline{C}_p|$\\
                \midrule
         GNN+SA & $\mathbf{81.2\pm 0.2}$ & $\mathbf{-0.004 \pm 0.029}$ &$\mathbf{0.026}$\\
         cVAE & $80.5\pm 0.2$ & $0.037 \pm 0.037$ & 0.089 \\
         \bottomrule
    \end{tabular}
    \vspace{10pt}
    \caption{Performance metrics comparing cardinality prediction accuracy, the MMD metric, and the difference in the mean Hungarian Cost between reconstructed and truth events of the surrogate models to the ground truth.}
    \label{tab:metrics}
\end{table*}
In Figure~\ref{fig:conditional_cardinality}, we compare the learned cardinality distributions as a function of the truth cardinality $p(N_R|N_T)$. Both models broadly reproduce the target with a slightly better performance achieved by  GNN+SA, as seen from table~\ref{tab:metrics}.
In Figure~\ref{fig:conditional_mean}, we compare the per-particle mean reconstructed momentum in bins of truth-momentum. In the ground truth reconstruction, the distribution of reconstructed momenta is as expected, centred around the true momenta, with the width reflecting the variance of the noise model and a residual variance contributed to the finite size bin-width in the conditional feature $\mathbf{t}$. Here, we can see that while the GNN+SA model does not fully match the ground truth, it performs markedly better than the cVAE model. The high variance of the cVAE model indicates that it does not model the mean of the reconstructed particle distributions correctly.
In Figure~\ref{fig:conditional_var}, the same analysis is performed for the variance of the reconstructed feature distribution and compared to the underlying ground-truth noise model that was applied to the truth particles. Here, the difference between the two models becomes even more apparent: Whereas the GNN+SA model does track the ground truth model, albeit with a degree of under-estimation and increased variance at high momenta, the cVAE model does not manage to correctly capture the momentum dependence of the resolution to a satisfying degree. The failures of the cVAE model are apparent in the representative truth event shown in Figure~\ref{fig:event_display}.
Comparing the results in Figure~\ref{fig:1d_pt} and Figures \ref{fig:conditional_distributions} underlines the importance of a detailed study of the learned set-valued distribution -- which is first done in this work -- as mismodelling may not be apparent from marginal distributions alone. Finally, we present a metric that aims to distil the interplay between multi-particle correlation as well as the shifts in the mean and variance modeling observed in the previous section into a single number. 
In Figure~\ref{fig:hungarian_cost_distr}, we show the distribution of the Hungarian Loss  $C(R,T)$ of the reconstructed events to the truth events as averaged over the full test set. The mean GNN+SA cost lies much closer to the mean ground-truth cost as compared to the cVAE model. To give a sense of scale for the significant improvement, a single event is shown in the main panel of Figure~\ref{fig:event_display}. The cVAE fails to sample reconstructed particles correctly, yielding a high Hungarian Cost. The GNN+SA samples resemble the target to a markedly higher degree. The cost distributions for this single truth event are shown as filled histograms in the inset. They each lie in the bulk of the truth-averaged distribution. The shown event is thus a representative example of the model performance. Similarly, the GNN+SA exceeds the cVAE performance as measured by $\mathrm{MMD}^2$ metric as listed in Table~\ref{tab:metrics}.
\section{Conclusions}
We have presented an approach for a set-conditional set generation model to approximate simulation and subsequent reconstruction.  We split the task into a two-step generative procedure of cardinality prediction followed by conditional set generation and choose appropriate permutation-invariant architectures through message-passing graph neural networks and slot-attention (GNN+SA). Results are shown on the reconstructions of the local collection of noised truth particles and compared to a baseline model that uses a cVAE architecture. The GNN+SA model outperforms the baseline model and better captures key properties of the target distribution.
While the current results may not be immediately applicable to real physics scenarios, our study demonstrates the feasibility of generating reconstructed particle sets while accurately modeling detector resolution using the proposed techniques and evaluation metrics. These findings are a foundation for further research in this area.

\noindent The main advantage of the current model is the joint simulation of sets of truth particles that exhibit non-trivial conditional dependencies in reconstruction efficiencies, e.g., sets of particles within spatially dense jets. On the other hand, the reconstruction of particles in disconnected regions is conditionally independent. Consequently, a viable path of scaling this approach to a fast simulation of full events is first to partition the full event truth particle set into groups that are assumed to be conditionally independent and apply a future version of the present model independently (and in parallel) to such clusters. The model must, therefore, only scale up to the effective partition size within an event and not to the full event cardinality itself.
For our future work, we intend to test this setup on a realistic data set that uses full detector simulation and reconstruction, includes neutral particles, and predicts each particle's class (i.e., whether it is a charged hadron, neutral hadron, electron, photon, or muon). We aim to improve accuracy and performance by investigating new architectures.

\section*{Acknowledgement}

\noindent The authors would like to thank Kyle Cranmer for the fruitful discussion and comments on the manuscript. ED is supported by the Zuckerman STEM Leadership Program. SG is partially supported by the Institute of AI and Beyond for the University of Tokyo. EG is supported by the Israel Science Foundation (ISF), Grant No. 2871/19 Centers of Excellence. LH is supported by the Excellence Cluster ORIGINS, which is funded by the Deutsche Forschungsgemeinschaft (DFG, German Research Foundation) under Germany’s Excellence Strategy - EXC-2094-390783311.

\newpage

\bibliography{fastsim}
\bibliographystyle{unsrt}

\end{document}